\documentstyle[aps,prb,epsf]{revtex}

\begin{document}

\renewcommand\baselinestretch{1.6}
\large\normalsize

\title{Dynamics of $zz$ spin correlations
       \protect\\
       in the square--lattice spin--$\frac{1}{2}$ isotropic $XY$ model}

\author{Oleg Derzhko$^{\dagger,\ddagger}$
and Taras Krokhmalskii$^{\dagger}$\\
{\small {$^{\dagger}$Institute for Condensed Matter Physics,
National Academy of Sciences of Ukraine}}\\
{\small {1 Svientsitskii Street, L'viv--11, 79011, Ukraine}}\\
{\small {$^{\ddagger}$Chair of Theoretical Physics,
Ivan Franko National University of L'viv}}\\
{\small {12 Drahomanov Street, L'viv--5, 79005, Ukraine}}}

\date{\today}

\maketitle

\begin{abstract}
Using the Jordan--Wigner fermionization in two dimensions
we obtain
the $zz$ wave vector-- and frequency--dependent structure factor
for the spin--$\frac{1}{2}$ isotropic $XY$ model
on a spatially anisotropic square lattice.
We use the obtained results
to discuss a role of the interchain interaction
for the dynamic properties of quasi--one--dimensional systems.
\end{abstract}

\vspace{5mm}


\vspace{5mm}


\vspace{5mm}

\noindent
{\bf {Postal addresses:}}\\
Dr. Oleg Derzhko (corresponding author)\\
Dr. Taras Krokhmalskii\\
Institute for Condensed Matter Physics, National Academy of Sciences of Ukraine\\
1 Svientsitskii Street, L'viv--11, 79011, Ukraine\\
tel: (0322) 76 19 78\\
fax: (0322) 76 11 58\\
email: derzhko@icmp.lviv.ua

\hspace{6mm}
krokhm@icmp.lviv.ua\\

\renewcommand\baselinestretch{2.2}
\large\normalsize

Dynamic properties 
of the two--dimensional quantum spin models
have attracted much attention for the last years
mainly due to the discovery 
of high--temperature superconductivity 
of the layered copper oxides \cite{001}.
At present, many compounds are known 
to be good realizations 
of the square--lattice quantum spin model 
and the experimental investigations of their dynamic properties \cite{002} 
require corresponding theoretical studies.
In most cases 
the isotropic Heisenberg Hamiltonian 
is used to appropriately model 
the spin degrees of freedom.
However, some generic features of the square--lattice quantum spin systems  
can be illustrated within the context 
of the spin--$\frac{1}{2}$ isotropic $XY$ model,
which is far more amenable to a theoretical analysis. 
Moreover, a study of the latter model 
can be viewed as a first step 
for examining the more complicated case 
of the square--lattice spin--$\frac{1}{2}$ Heisenberg model. 

The aim of this study is to examine the dynamic properties 
of the spin--$\frac{1}{2}$ isotropic $XY$ model on a square lattice. 
Many theoretical and numerical investigations 
concerning the thermodynamic properties of this model 
have been performed until now \cite{003}.
Less progress has been made in the studies 
of its dynamic properties.
To examine the dynamic properties 
of the model 
we use the two--dimensional Jordan--Wigner fermionization \cite{004,005,006,007}
(for a brief review see Ref. \onlinecite{008})
thus extending the approach 
which was previously applied mainly to the study of thermodynamics
(see also several recent papers \cite{008a} 
on calculation of the magnetization curves 
of some two--dimensional spin systems).
More specifically,
we calculate 
the two--site time--dependent correlation functions of $z$ spin components
and the corresponding dynamic structure factor 
$S_{zz}({\bf{k}},\omega)$
for the spin--$\frac{1}{2}$ isotropic $XY$ model 
on a spatially anisotropic square lattice.
To trace a one--dimensional to two--dimensional crossover,
we compare the results obtained 
with the corresponding ones 
for the spin--$\frac{1}{2}$ isotropic $XY$ chain 
thus 
demonstrating the role of the interchain interaction 
for the dynamic properties 
of quasi--one--dimensional spin--$\frac{1}{2}$ isotropic $XY$ systems.
The recent study of the static properties 
of the square--lattice spin--$\frac{1}{2}$ isotropic $XY$ model \cite{008b} 
suggests 
that the approach based on the two--dimensional Jordan--Wigner fermionization works well 
as long as the interchain interaction is small 
and it gives qualitatively correct results 
for larger values of the interchain interaction. 

We consider 
a spin model consisting of $N\to\infty$ spins $\frac{1}{2}$ 
on a spatially anisotropic square lattice
governed by the isotropic $XY$ Hamiltonian
\begin{eqnarray}
\label{001}
H=\frac{1}{2}\sum_{i=0}^{\infty}\sum_{j=0}^{\infty}
\left(
J\left(s^+_{i,j} s^-_{i+1,j}+s^-_{i,j} s^+_{i+1,j}\right)
+
J_{\perp}\left(s^+_{i,j} s^-_{i,j+1}+s^-_{i,j} s^+_{i,j+1}\right)
\right).
\end{eqnarray}
Here $J\ge 0$ and $J_{\perp}\ge 0$ 
are the exchange interactions between the neighbouring sites
in a row and in a column, respectively.
We are interested 
in the two--site time--dependent spin correlation functions 
$\langle s^z_{n,m}(t)s^z_{n+p,m+q} \rangle$ 
(the angular brackets denote the canonical thermodynamic average 
with the Hamiltonian (\ref{001})) 
and the $zz$ dynamic structure factor 
\begin{eqnarray}
\label{002}
S_{zz}({\bf{k}},\omega)
=
\sum_{p=0}^{\infty}\sum_{q=0}^{\infty}
{\mbox{e}}^{{\mbox{i}}\left(k_xp+k_yq\right)}
\int_{-\infty}^{\infty}{\mbox{d}}t
{\mbox{e}}^{{\mbox{i}}\omega t}
\left(
\langle s^z_{n,m}(t)s^z_{n+p,m+q} \rangle
-\langle s^z_{n,m} \rangle \langle s^z_{n+p,m+q} \rangle 
\right).
\end{eqnarray}

To calculate these quantities 
we use the two--dimensional Jordan--Wigner transformation \cite{005} 
reformulating the problem in fermionic language. 
First we introduce
the annihilation and creation operators of (spinless) fermion
$d_{i,j}
={\mbox{e}}^{-{\mbox{i}}\alpha_{i,j}}s_{i,j}^-$,
$\alpha_{i,j}
=\sum_{f=0(\ne i)}^{\infty}
\sum_{g=0(\ne j)}^{\infty}
B_{i,j;f,g}d^+_{f,g}d_{f,g}$,
$B_{i,j;f,g}
={\mbox{Im}}\ln\left(f-i+{\mbox{i}}\left(g-j\right)\right)$.
We adopt the mean--field treatment of the phase factors 
which appear in the initial Hamiltonian (\ref{001}) 
after fermionization 
(for details see Refs. \onlinecite{005,008}). 
This is the only approximation made.
Then we perform 
the Fourier transformation
and 
the Bogolyubov transformation
to arrive at the Hamiltonian of noninteracting fermions.
Next, 
we rewrite the correlation function 
$\langle s^z_{n,m}(t)s^z_{n+p,m+q} \rangle$
in fermionic language
and use the Wick--Bloch--de Dominicis theorem to calculate it.
Finally, 
we insert the derived result into (\ref{002}) 
and get the $zz$ dynamic structure factor
\begin{eqnarray}
\label{003}
S_{zz}({\bf{k}},\omega)=
\pi
\int_{-\pi}^{\pi}\frac{{\mbox{d}}k_{1y}}{2\pi}
\int_{-\pi}^{\pi}\frac{{\mbox{d}}k_{1x}}{2\pi}
\nonumber\\
\left(
\cos^2\frac{\gamma_{{\bf{k}}_1+{\bf{k}}}-\gamma_{{\bf{k}}_1}}{2}
\left(n_{{\bf{k}}_1}
\left(1-n_{{\bf{k}}_1+{\bf{k}}}\right)
\delta\left(\omega+\Lambda_{{\bf{k}}_1}-
\Lambda_{{\bf{k}}_1+{\bf{k}}}\right)
\right.
\right.
\nonumber\\
\left.
\left.
+
\left(1-n_{{\bf{k}}_1}\right)n_{{\bf{k}}_1+{\bf{k}}}
\delta\left(\omega-\Lambda_{{\bf{k}}_1}+
\Lambda_{{\bf{k}}_1+{\bf{k}}}\right)
\right)
\right.
\nonumber\\
\left.
+\sin^2\frac{\gamma_{{\bf{k}}_1+{\bf{k}}}-\gamma_{{\bf{k}}_1}}{2}
\left(n_{{\bf{k}}_1}n_{{\bf{k}}_1+{\bf{k}}}
\delta\left(\omega+\Lambda_{{\bf{k}}_1}+
\Lambda_{{\bf{k}}_1+{\bf{k}}}\right)
\right.
\right.
\nonumber\\
\left.
\left.
+
\left(1-n_{{\bf{k}}_1}\right)
\left(1-n_{{\bf{k}}_1+{\bf{k}}}\right)
\delta\left(\omega-\Lambda_{{\bf{k}}_1}-
\Lambda_{{\bf{k}}_1+{\bf{k}}}\right)
\right)
\right).
\end{eqnarray}
Here
$\Lambda_{{\bf{k}}}
=\sqrt{J^2\sin^2k_x+J_{\perp}^2\cos^2k_y}\ge 0$
is the fermion energy,
$\cos\frac{\gamma_{{\bf{k}}}}{2}
=\sqrt{\frac{1}{2}
+\frac{J_{\perp}\cos k_y}{2\Lambda_{{\bf{k}}}}}$,
$\sin\frac{\gamma_{{\bf{k}}}}{2}
={\mbox{sgn}}\left(J\sin k_x\right)
\sqrt{\frac{1}{2}
-\frac{J_{\perp}\cos k_y}{2\Lambda_{{\bf{k}}}}}$,
and 
$n_{\bf{k}}=\frac{1}{{\mbox{e}}^{\beta\Lambda_{\bf{k}}}+1}$
is the Fermi factor.
In Figs. 1 -- 3 
we present $S_{zz}({\bf{k}},\omega)$ 
for the square--lattice spin--$\frac{1}{2}$ isotropic $XY$ model
as it follows from Eq. (\ref{003})
(to integrate over $k_{1x}$
we use the relation 
$\delta\left(\varphi(x)\right)
=\sum_j\delta\left(x-a_j\right)
\vert \varphi^{\prime}(a_j)\vert^{-1}$,
$a_j$ are the roots of equation 
$\varphi(x)=0$,
and then we do the integral over $k_{1y}$ numerically).  
Let us discuss 
the dynamic properties of the considered spin model (\ref{001})
concentrating in particular 
on the appearance of one--dimensional behaviour  
at low and high temperatures
as $\frac{J_{\perp}}{J}$ decreases.

It is useful to start the analysis of dynamic properties 
by reminding what is long known for the one--dimensional case.
From the exact calculation 
for spin--$\frac{1}{2}$ isotropic $XY$ chain \cite{009}
we know 
that $S_{zz}(k,\omega)$ at zero temperature $\beta=\infty$
is governed by a two--fermion excitation continuum
with lower and upper boundaries
$J\vert\sin k\vert$
and
$2J\vert\sin\frac{k}{2}\vert$,
respectively.
The spectral weight is finite at the lower boundary.
The lower boundary touches the line $\omega=0$ 
at $k=0,\;2\pi$ and $k=\pi$
(gapless modes).
The spectral weight increases towards the upper boundary 
where $S_{zz}(k,\omega)$ diverges.
With the increasing temperature, 
the lower boundary gets smeared out and finally disappears,
whereas the upper boundary remains unchanged.
The described picture agrees with what we see 
from the greyscale plots of $S_{zz}(k_x,0,\omega)$
for $\frac{J_{\perp}}{J}=0.1$
(Figs. 1a and 2a),
i.e., for a system of almost noninteracting chains 
for momentum transfer along the chain direction.
(The gap seen in Fig. 2a arises due to a finite interchain interaction:
as $J_{\perp}\to 0$ it shrinks and disappears.)
For momentum transfer perpendicularly to the chain direction, 
the dynamic structure factor 
for a system of almost noninteracting chains 
disappears 
(Figs. 1b and 2b).

Let us pass to the two--dimensional case
(the rest of the plots in Figs. 1, 2).
Similar to the one--dimensional case   
$S_{zz}({\bf{k}},\omega)$
within the framework of our approach 
is conditioned by two--fermion excitations 
and hence it exhibits nonzero values 
only within the restricted region 
having a sharp high--frequency cutoff
(evidently,
$S_{zz}({\bf{k}},\omega)$ (\ref{003})
equals to zero 
if the frequency $\omega$ exceeds $2\sqrt{J^2+J_{\perp}^2}$).
Contrary to the one--dimensional case, 
for which $S_{zz}(k,\omega)$ is almost structureless
within the excitation continuum
(apart from the upper boundary singularities),
$S_{zz}({\bf{k}},\omega)$ 
exhibits several washed--out excitation branches (modes).
Consider at first zero temperature.
As it follows from Eq. (\ref{003}), 
the $\delta$--function 
$\delta\left(\omega-\Lambda_{{\bf{k}}_1}
-\Lambda_{{\bf{k}}_1+{\bf{k}}}\right)$ 
admits the spectral weight at $\bf{k},\omega$ 
due to the two fermions with
${\bf{k}}_1
=\left(-\frac{k_x}{2},\frac{\pi}{2}-\frac{k_y}{2}\right)$
and 
${\bf{k}}_1+{\bf{k}}
=\left(\frac{k_x}{2},\frac{\pi}{2}+\frac{k_y}{2}\right)$
with the energy of the pair
\begin{eqnarray}
\label{004}
\omega^{(1)}_{{\bf{k}}}=
2\sqrt{J^2\sin^2\frac{k_x}{2}+J^2_{\perp}\sin^2\frac{k_y}{2}}.
\end{eqnarray}
This excitation is known as the spin wave \cite{003}.
Note,
that since the derivative of the argument of $\delta$--function 
with respect to $k_{1x}$ tends to zero 
(except for $k_x=0,\;2\pi$ if $k_y=0,\;2\pi$ and vise versa)
the peaks of $S_{zz}({\bf{k}},\omega)$ 
along the line $\omega^{(1)}_{{\bf{k}}}$ (\ref{004}) 
are expected.
They are nicely visible in Figs. 1, 2.

It is easy to note
that the branch 
which almost everywhere 
in the regions 
$\left(k_x, 0, \omega\right)$ 
and 
$\left(0, k_y, \omega\right)$
forms the upper boundary 
(see Figs. 1, 2)
is given by
\begin{eqnarray}
\label{005}
\omega^{(2)}_{k_x,0}=2\sqrt{J^2\sin^2\frac{k_x}{2}+J^2_{\perp}},
\;\;\;
\omega^{(2)}_{0,k_y}=2\sqrt{J^2+J^2_{\perp}\sin^2\frac{k_y}{2}}.
\end{eqnarray}
$S_{zz}({\bf{k}},\omega)$ arises along the lines 
$\omega^{(2)}_{k_x,0}$ and $\omega^{(2)}_{0,k_y}$
(\ref{005}) 
when it is conditioned by the two fermions with
$\left(-\frac{k_x}{2},0\right)$,
$\left(\frac{k_x}{2},0\right)$
and 
$\left(\frac{\pi}{2},\frac{\pi}{2}-\frac{k_y}{2}\right)$,
$\left(\frac{\pi}{2},\frac{\pi}{2}+\frac{k_y}{2}\right)$,
respectively.
There are other high--frequency excitations 
in Figs. 1, 2,
conditioned by two fermions  
with
${\bf{k}}_1=\left(0,0\right)$ 
and
${\bf{k}}_1+{\bf{k}}$
($k_y=0$)
or
with
${\bf{k}}_1=\left(\frac{\pi}{2},\frac{\pi}{2}\right)$
and
${\bf{k}}_1+{\bf{k}}$
($k_x=0$),
i.e.,
\begin{eqnarray}
\label{006}
\omega^{(3)}_{k_x,0}=J_{\perp}+\sqrt{J^2\sin^2 k_x+J^2_{\perp}},
\;\;\;
\omega^{(3)}_{0,k_y}=J+\sqrt{J^2+J^2_{\perp}\sin^2 k_y}.
\end{eqnarray}
It is interesting to note, 
that the derivative of the argument of
$\delta\left(\omega-\Lambda_{{\bf{k}}_1}
-\Lambda_{{\bf{k}}_1+{\bf{k}}}\right)$ 
with respect to $k_{1x}$ 
tends to zero 
at $k_x=\frac{\pi}{2},\;\frac{3\pi}{2}$ ($k_y=0$)
or 
at $k_y=\frac{\pi}{2},\;\frac{3\pi}{2}$ ($k_x=0$)
that manifests itself 
as the increase of spectral weight in these regions
clearly seen in some plots in Figs. 1, 2.

In the low--temperature limit, 
because of the Fermi factors, 
the $\delta$--functions 
$\delta\left(\omega-\Lambda_{{\bf{k}}_1}
+\Lambda_{{\bf{k}}_1+{\bf{k}}}\right)$ 
and
$\delta\left(\omega+\Lambda_{{\bf{k}}_1}
-\Lambda_{{\bf{k}}_1+{\bf{k}}}\right)$ 
in Eq. (\ref{003})
may contribute only 
if the energy of one of the two fermions equals to zero,
i.e.,
if either
$\Lambda_{{\bf{k}}_1}=0$
or 
$\Lambda_{{\bf{k}}_1+{\bf{k}}}=0$.
As a result, 
we easily derive the lower boundary 
for the region of nonzero values of $S_{zz}({\bf{k}},\omega)$
\begin{eqnarray}
\label{007}
\omega^{(4)}_{{\bf{k}}}=
\sqrt{J^2\sin^2 k_x+J^2_{\perp}\sin^2 k_y}.
\end{eqnarray}
At high temperatures, 
the Fermi factors do not impose the mentioned restriction.
Putting $k_{1x}=-k_x$, $k_{1y}=\frac{\pi}{2}-k_x$ for $k_y=0$ 
and $k_{1x}=k_y$, $k_{1y}=\frac{\pi}{2}-k_y$ for $k_x=0$
we get 
\begin{eqnarray}
\label{008}
\omega^{(5)}_{k_x,0}
=\left(\sqrt{J^2+J^2_{\perp}}-J_{\perp}\right)\vert\sin k_x\vert,
\;\;\;
\omega^{(5)}_{0,k_y}
=\left(\sqrt{J^2+J^2_{\perp}}-J\right)\vert\sin k_y\vert.
\end{eqnarray}
The excitation branch $\omega^{(5)}_{{\bf{k}}}$ (\ref{008}) 
contains most of the spectral weight at high temperatures 
(see Fig. 2).

Obviously in our analysis of $S_{zz}({\bf{k}},\omega)$ we focus mainly 
on the consequences 
presumed by the $\delta$--functions involved in (\ref{003}) 
omitting the precise examination of a role of the rest of the factors 
(e.g., of the Fermi factors 
which controls a redistribution of the spectral weight 
as temperature increases).
On the other hand,
the results presented in Figs. 1, 2 
(and Fig. 3)
reflect all those underlying features of Eq. (\ref{003}).

In Fig. 3, we show the constant wave vector scans 
of $zz$ dynamic structure factor,
i.e., the frequency dependence of 
$S_{zz}({\bf{k}},\omega)$ 
for certain wave vectors 
(the corresponding sections in Figs. 1, 2 
can be easily located).
All the discussed properties of $S_{zz}({\bf{k}},\omega)$
clearly manifest themselves 
in the depicted frequency profiles 
which may be almost symmetric or asymmetric,
resemble $\delta$--peaks,
result from two coalesced peaks
or exhibit tails
which gradually disappear or are abruptly cut off.
We indicate by numbers $j$ some well pronounced peaks and cusps
corresponding to the modes $\omega_{\bf{k}}^{(j)}$
(\ref{004}) -- ({\ref{008}).

To summarize,
we have presented the first results
for $zz$ dynamic structure factor
of the spin--$\frac{1}{2}$ isotropic $XY$ model
on a spatially anisotropic square lattice
derived on the basis
of the two--dimensional Jordan--Wigner fermionization.
Our study has indicated excitations
which govern the $zz$ dynamic structure factor.
The established modes
may manifest themselves as peaks or cusps
in frequency profiles
of $S_{zz}({\bf{k}},\omega)$
and may be used for determining the Hamiltonian parameters.
(For example,
the gap seen in Fig. 2a and Fig. 3e
is controlled by
$\omega^{(3)}_{k_x,0}$,
$\omega^{(4)}_{k_x,0}$,
and
$\omega^{(5)}_{k_x,0}$,
and, in particular,
$\omega^{(3)}_{\frac{\pi}{2},0}=\sqrt{J^2+J^2_{\perp}}+J_{\perp}$,
$\omega^{(4)}_{\frac{\pi}{2},0}=J$,
and
$\omega^{(5)}_{\frac{\pi}{2},0}=\sqrt{J^2+J^2_{\perp}}-J_{\perp}$.)
Thus,
the theoretical result observed in our work
should prove valuable
in understanding the experimentally observable dynamic properties
of two--dimensional spin--$\frac{1}{2}$ isotropic $XY$ materials.
However,
there are, to our knowledge,
no experimental results yet available
(therefore,
no compound specific parameters were considered above)
that enable a direct comparison between theory and experiment
to be made.

It should be stressed
that our results for the considered two--dimensional spin model
contains the exact ones in the one--dimensional limit
(after putting $J_{\perp}=0$ or $J=0$).
On the other hand,
our approach is approximate
because of the mean--field description
of the phase factors
which arise after fermionization.
Such a treatment neglects
the complicated interaction between fermions.
It would be desirable
to estimate the effects of this simplification
going beyond the mean--field scheme for the phase factors.
A possible test of the results for spin correlation functions
is a comparison
with the corresponding data of exact diagonalization and other numerical approaches.
(Note,
however,
that the results reported in Refs. \onlinecite{003,008b}
refer to the in--plane spin correlation functions
but not to the $zz$ spin correlation functions
which are different for the isotropic $XY$ model
in contrast to the case of the isotropic Heisenberg model.)
In the case of the isotropic Heisenberg model
the interaction between fermions is present
even within the adopted approximate procedure
due to the (Ising) interaction of $z$ spin components.
The quartic terms in the fermion Hamiltonian
are treated after making further approximation \cite{005}.
One of the ways to proceed
(the uniform flux solution)
yields the same results for both the $XY$ and Heisenberg models
and from such a viewpoint
our results for the dynamic properties
may refer to
the square--lattice spin--$\frac{1}{2}$ isotropic Heisenberg model
as well.
However,
as one can conclude from the studies
reported in Refs. \onlinecite{005,006},
a more sophisticated treatment
of the fermion interaction
inherent in the latter spin model
(the in--phase flux solution,
the N\'{e}el flux solution,
the in--phase N\'{e}el flux solution)
would be more appropriate.
It seems obvious to extend the present investigation
to the study
of the square--lattice spin--$\frac{1}{2}$ isotropic Heisenberg model
which is used for interpreting
the data for corresponding compounds \cite{002}.
It is interesting to note
that the recent neutron scattering studies on Cs$_2$CuCl$_4$
(the minimal model of this two--dimensional compound
is the spin--$\frac{1}{2}$ antiferromagnetic Heisenberg model
on a triangular lattice) \cite{009a}
definitely show a highly dispersive continuum of excited states.
The conventional spin--wave theory clearly fails
to account for the observed line shapes
which do not show single particle poles
but rather extended continua.
We can also see some similarities at a qualitative level
of the frequency profiles
plotted in Fig. 3 of the present paper
and in Fig. 3 of Ref. \onlinecite{009a}.
However,
we should bear in mind
that Cs$_2$CuCl$_4$ is a triangular but not a square--lattice two--dimensional system
and
that the minimal Hamiltonian determining the magnetic order is the isotropic Heisenberg model
but not the isotropic $XY$ model.
New theoretical work is needed
to elaborate the two--dimensional Jordan--Wigner fermionization results for this compound.
Another possible application
refers to dynamics of spin--$\frac{1}{2}$ ladders.
Recently  \cite{010}
the dynamic properties of the two--leg spin--$\frac{1}{2}$ ladder
have been examined using the one--dimensional Jordan--Wigner fermionization.
It seems interesting to compare these results
with the corresponding ones
which arise within the two--dimensional Jordan--Wigner fermionization approach.

The authors thank S. Haas
for a critical reading of the manuscript
and helpful comments.
O. D. is grateful to the Max--Planck--Institut f\"{u}r Physik komplexer Systeme (Dresden)
for the hospitality during the International Workshop and Seminar
on Modern Aspects of Quantum Impurity Physics (2003)
when the paper was completed.

\vspace{25mm}

FIGURE CAPTURES

\vspace{2mm}

FIGURE 1.
The $zz$ dynamic structure factor 
(greyscale plots)
for the square--lattice spin--$\frac{1}{2}$ isotropic $XY$ model 
as it follows from Eq. (\ref{003}) 
at zero temperature.
$J=1$, 
$J_{\perp}=0.1$ (a, b),
$J_{\perp}=0.5$ (c, d),
$J_{\perp}=0.9$ (e, f);
$k_{y}=0$ (a, c, e),
$k_{x}=0$ (b, d, f).

\vspace{2mm}

FIGURE 2.
The same as in Fig. 2
at high temperature 
$\beta=0.1$.

\vspace{2mm}

FIGURE 3.
Frequency dependence of the $zz$ dynamic structure factor 
(\ref{003})
for momentum transfer along the chain 
($k_x=\frac{\pi}{2}$ (a, e),
$k_x=\pi$ (b, f),
$k_y=0$)
and 
for momentum transfer perpendicularly to the chain
($k_x=0$,
$k_y=\frac{\pi}{2}$ (c),
$k_y=\pi$ (d))
as the interchain interaction changes
($J=1$,
$J_{\perp}=0.1$ (dotted curves),
$J_{\perp}=0.5$ (dashed curves),
$J_{\perp}=0.9$ (solid curves))
at zero temperature (a, b, c, d)
and
high temperature
$\beta=0.1$ (e, f).

\end{document}